\documentclass[pra, aps]{revtex4}
\usepackage{graphicx}
\usepackage{amssymb}
\usepackage{xcolor}
\usepackage{hyperref}

\newcommand{\Tr}{{\rm{Tr}}}
\begin{document}

\title{Theoretical investigations of quantum correlations in NMR multiple-pulse spin-locking experiments
}


\author{S. A. Gerasev $^{1, 2}$, A. V. Fedorova$^{1}$, E.~B.~Fel'dman$^{1}$, E. I. Kuznetsova$^{1}$, }
\affiliation{$^1$ Institute of Problems of Chemical Physics, Chernogolovka, Moscow Region,
142432, RUSSIA}
\affiliation{$^2$ 119991, Moscow, Lomonosov Moscow State University, Faculty of Physics and Chemistry}


\begin{abstract}
Quantum correlations are investigated theoretically in a two-spin system with the dipole-dipole interactions in the NMR multiple-pulse spin-locking experiments. We consider two schemes of the multiple-pulse spin-locking. The first scheme consists of   $\pi/2$-pulses only and the delays between the pulses can differ. The second scheme contains $\varphi$-pulses ($0<\varphi<\pi$) and has equal delays between them. 
We calculate entanglement for both schemes for an initial separable state.
We show that entanglement is absent for the first scheme at equal delays between $\pi/2$-pulses at arbitraty temperatures. 
Entanglement emerges after several periods of the pulse sequence in the second scheme at $\varphi=\pi/4$ at milliKelvin temperatures.
The necessary number of the periods increases with increasing temperature. We demonstrate the dependence of entanglement on the number of the periods of the multiple-pulse sequence.
Quantum discord is obtained for the first scheme of the multiple-pulse spin-locking experiment at different temperatures.
\end{abstract}

\maketitle

\section{Introduction}
\label{intro}
Multiple pulse NMR spectroscopy allowed us to obtain NMR spectra of high resolution and to investigate slow relaxation processes in solids \cite{haeberlen}. The multiple-pulse 
spin-locking was one of the first multiple-pulse NMR experiments \cite{ostroff,mansfield}.
In those experiments the initial thermodynamic equilibrium magnetization 
along axis $z$ %÷òî-òî ìíå íå íðàâèòñÿ ÷èñëî îñåé
is turned  $90^\circ$ into the perpendicular plane (axis $x$) by a resonance~$90^\circ_y$ pulse.
Then the periodic sequence of  r.f. resonance $x$-pulses locks the magnetization along axis $x$. Measurements of the magnetization in the windows between the pulses yield information about relaxation processes in solids \cite{ostroff,mansfield}. It was shown experimentally that the multiple-pulse spin-locking experiment with the $\pi/2$-resonance pulses performs a long chain of echo signals with the amplitudes decaying on the time scale of $T_{2}\sim T_2/(\omega_{loc}\tau)^4$ \cite{D1,D2}, where $T_2$ is the time of spin-spin relaxation, $\omega_{loc}$ is the local dipolar frequency, and $\tau$ is proportional to the delay between two successive pulses. The theoretical confirmation of this result is given in \cite{D3}. The multiple-pulse spin-locking experiment brings a significant advantage in time  in comparison with the continuous spin locking experiment at relaxation measurements.

It has been noted recently \cite{kuznets,feldman} that the multiple-pulse spin-locking experiment can be used for the investigation of quantum correlations. 
In fact, this experiment can be seen as a simple tool for the study of quantum correlations in a system subject to an external periodic field.
It is very important that the decoherence time in the multi-pulse spin locking experiment is determined by the spin-lattice relaxation times in the rotating reference frame \cite{goldman}. These times are usually very long (more than 10 s).
Starting with a system in a separable state, it is possible to show how entangled states emerge in the multiple spin-locking after several periods of the external periodic perturbation.
The conditions for the emergence of quantum correlations in the multiple-pulse spin-locking can be also found.
The main goal of this paper is an investigation of quantum correlations in a two-spin system with the dipole-dipole interaction (DDI) in the multiple-pulse spin-locking on the basis of calculations of entanglement \cite{hill,wootters} and quantum discord \cite{ollivier,henderson}.

The paper is organized as follows. 
In Section 2, we describe two schemes of the multiple-pulse spin-locking which will be used for investigations of quantum correlations.
In Section 3, we study entanglement in a two-spin system with the DDI in both schemes. 
In Section 4, we investigate quantum discord for the first scheme of the multiple-pulse spin-locking at low and high temperatures.
In particular, analytical calculations of the quantum discord are performed in the high temperature approximation \cite{goldman}.
We briefly summarize our results in Section 5.

\section{Two schemes of the multiple-pulse spin-locking NMR experiment}
We consider a two-spin system with the DDI in a strong external magnetic field directed along the $z$-axis of the laboratory reference frame.
The secular part of the DDI with respect to the external magnetic field  $\vec{H_0}$ can be written as follows \cite{goldman}:
\begin{equation}
H_{dz}=\sum_{i<j} d_{ij} (3 I_{iz} I_{jz} - \vec{I_i}\cdot \vec{I_j}),
\end{equation}
where $I_{i\alpha}$ ($\alpha=x,y,z$) is the projection of the angular momentum of spin $i$ on axis $\alpha$, $\vec{I_i}\cdot \vec{I_j} =I_{ix} I_{jx} +I_{iy} I_{jy} +I_{iz} I_{jz}$ and $d_{ij}$ is the dipolar coupling constant of spins $i$ and $j$.

\begin{figure}
  \includegraphics[scale=0.3]{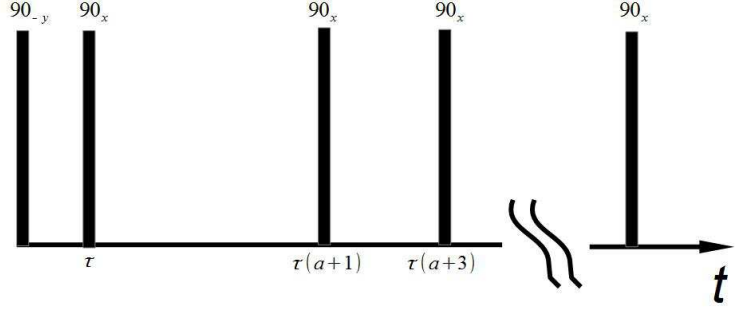}
\caption{Scheme $A$ of the pulse sequence of the multiple-pulse spin-locking experiment; $90_{-y}$ is the initial pulse turning the magnetization from the  $z$ axis to the $x$-axis; $90_x$ is the pulse turning the magnetization about axis $x$ by the angle $90^\circ$; $\tau$, $a\tau$, and $2\tau$ are the delays between two successive pulses; $a$ is a positive number.}
\label{fig:1}       % Give a unique label
\end{figure}

The multiple-pulse spin-locking experiment consists of a sequence of resonance pulses of the magnetic  field. It is convenient to describe their effect in the rotating reference frame (RRF) \cite{goldman} which rotates spins with the Larmor frequency $\omega_0=\gamma H_0$, where $\gamma$ is the gyromagnetic ratio. 
The multiple-pulse spin-locking experiment involves an initial $\pi/2$ resonance $y$-pulse that rotates spins by $90^\circ$ about the $y$-axis of the RRF.
We denote that pulse as $P^{90^\circ}_{-y}$.
In the first scheme (scheme A), after the delay time $\tau$, the multiple-pulse sequence of resonance pulses $P^{90^\circ}_x$, separated by alternating delays $a \tau$ and $2 \tau$ ($a$ is a positive number), is applied (Fig.~\ref{fig:1}).
Those pulses rotate spins by $90^\circ$ about the $x$ axis; $K$ is the total number of the $P^{90^\circ}_x$ pulses. For brevity, the set of the $K$ pulses is conventionally denoted as \cite{kuznets}
\begin{eqnarray}\label{eq2}
(P^{90^\circ}_x - a \tau &-& P^{90^\circ}_x - 2 \tau-{})^K= P^{90^\circ}_x - a\tau -P^{90^\circ}_x  - 2\tau \nonumber\\
{} &-& P^{90^\circ}_x - a \tau -P^{90^\circ}_x-2\tau \dots\dots P^{90^\circ}_x - a \tau -P^{90^\circ}_x-2\tau.
\end{eqnarray}
The total scheme $A$ of the multiple-pulse spin-locking experiment can be represented as
\begin{equation}\label{eq3}
P^{90^\circ}_{-y} -\tau-(P^{90^\circ}_x - a\tau - P^{90^\circ}_x - 2\tau - )^K.
\end{equation}
\begin{figure}
  \includegraphics[scale=0.3]{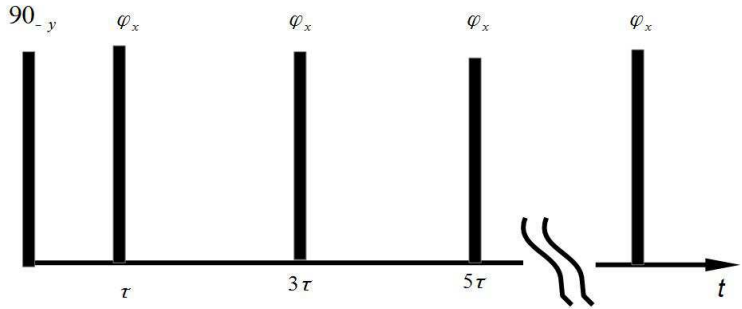}
\caption{Scheme $B$ of the pulse sequence of the multiple-pulse spin-locking experiment: $90_{-y}$ is the initial pulse turning the magnetization from the $z$ axis to the $x$-axis: $\varphi_x$ is the pulse turning the magnetization about axis  $x$ by the angle  $\varphi$; $2\tau$ is the delay between two successive pulses.}
\label{fig:2}       % Give a unique label
\end{figure}
In the second scheme (scheme B) the multiple-pulse sequence of resonance pulses $P^\varphi _x$, separated by delays $2\tau$, is applied (Fig. \ref{fig:2}). Those pulses rotate spins by the angle $\varphi$ about the $x$ axis. The total scheme B of the multiple spin locking experiment can be written as \cite{feldman}:
\begin{equation}\label{Eq4}
P^{90^\circ}_{-y} - \tau - (P^\varphi_x -2\tau-)^K.
\end{equation} 
We suppose that all pulses in Eqs.(\ref{eq2}), (\ref{eq3}), (\ref{Eq4}) are $\delta$-pulses. The Liouville equation for the density matrix $\rho(t)$ in the RRF is 
\begin{equation}
i\frac{d\rho}{dt}=\left[-f_S(t) I_x+H_{dz},\rho(t)\right],
\end{equation}
where $I_x=I_{1x}+I_{2x}$ and the pulse functions $f_S(t)$ ($S=A,\, B$) are
\begin{equation}
f_A(t)=-\frac{\pi}{2}\sum\limits^{[K]/2}_{j=0}\{\delta(\tau(a+1)+j(a+2)\tau-t)+\delta(\tau+j(a+2)\tau -t)\},
\end{equation}
\begin{equation}
f_B(t)=-\varphi\sum\limits^{K}_{j=0}\delta(\tau+2j\tau-t)
\end{equation}
for schemes $A$ and $B$, respectively.

\section{Entanglement in a two-spin system with the DDI in two schemes of the multiple-pulse NMR spin-locking}

Initially, the spin system is in the thermodynamic equilibrium state in the strong external magnetic field $\vec{H_0}$. After the initial preparation pulse $P^{90^\circ}_{-y}$, the density matrix $\rho_0$ becomes
\begin{equation}\label{dmatrix}
\rho_0=\frac{1}{Z}e^{\beta I_{1x}}\otimes e^{\beta I_{2x}},\qquad \beta=\frac{\hbar \omega_0}{k_B T},\qquad Z=4\cosh^2\frac{\beta}{2},
\end{equation}
where $T$ is the temperature and $Z$ is the partition function. It is evident that the state of Eq.~(\ref{dmatrix}) is separable. In scheme $A$ of the multiple-pulse spin-locking experiment at times $0<t<\tau$ (see Fig.~1), the DDI is
\begin{equation}\label{ham_dz}
H_{dz}=d(3I_{1z}I_{2z}-\vec{I_1}\cdot\vec{I_2}),
\end{equation}
where $d=d_{12}$.
After the first $\pi/2$-pulse, the anisotropic DDI can be defined as
\begin{equation}\label{ham_dy}
e^{-i\frac{\pi}{2}I_x}H_{dz}e^{i\frac{\pi}{2}I_x}=d(3I_{1y}I_{2y}-\vec{I_1}\cdot \vec{I_2})=H_{dy},
\end{equation}
where we take into account that
\begin{equation}\label{I_y}
e^{i\frac{\pi}{2}I_x}I_{iz}e^{-i\frac{\pi}{2}I_x}=e^{i\frac{\pi}{2}I_{ix}}I_{iz}e^{-i\frac{\pi}{2}I_{ix}}=I_{iy}, \qquad i=1,2
\end{equation}
and the scalar product $\vec{I_1}\cdot\vec{I_2}$ does not change with unitary transformations. The DDI is $H_{dy}$ in the interval $\tau<t<\tau(a+1)$.

We will follow a standard convention and label the spin-up and -down states as 1 and 0. The matrix representation of the Hamiltonians of Eqs.~(\ref{ham_dz}), (\ref{ham_dy}) in the basis  $|00\rangle$, $|01\rangle$, $|10\rangle$, $|11\rangle$ is \cite{nielsen}
\begin{equation}
H_{dz}=\frac{d}{2}\left(
\begin{array}{cccc}
1&0&0&0\\
0&-1&-1&0\\
0&-1&-1&0\\
0&0&0&1
\end{array}\right),
\qquad
H_{dy}=\frac{d}{4}
\left(
\begin{array}{cccc}
-1&0&0&-3\\
0&1&1&0\\
0&1&1&0\\
-3&0&0&-1
\end{array}
\right).
\end{equation}
The density matrix $\rho((a+2)\tau)$ at the end of the period of the sequence of scheme $A$ is
\begin{eqnarray}
\nonumber \rho((a+2)\tau)=e^{-i\tau H_{dz}}e^{-i\frac{\pi}{2} I_x}e^{-ia\tau H_{dz}} e^{-i\frac{\pi}{2}I_x}  e^{-i\tau H_{dz}}  \rho_0 e^{i \tau H_{dz}} e^{i\frac{\pi}{2}I_x}  e^{ia\tau H_{dz}}\\ 
{} \times e^{i\frac{\pi}{2} I_x} e^{i\tau H_{dz}}=
e^{-i\tau H_{dz}} e^{-ia\tau H_{dy}} e^{-i\tau H_{dz}}\rho_0 e^{i\tau H_{dz}} e^{ia\tau H_{dy}} e^{i\tau H_{dz}}.
\end{eqnarray}
We take into account that
\begin{equation}
e^{-i\frac{\pi}{2} I_x} H_{dz} e^{i \frac{\pi}{2} I_x}=H_{dy}, \qquad \left[e^{-i\pi I_x}, H_{dz}\right]=\left[e^{-i\pi I_x},\rho_0 \right]=0.
\end{equation}
Hence, the evolution operator $U((a+2)\tau)$ for one period of the sequence of scheme $A$ is
\begin{equation}\label{operator}
U((a+2)\tau)=e^{-i\tau H_{dz}}e^{-ia \tau H_{dy}} e ^{-i\tau H_{dz}}
\end{equation}
and in the matrix form
\begin{equation}\label{operator_matrix}
\begin{array}{l}
U((a+2)\tau)=\nonumber\\
\qquad\frac{1}{2}\left(
\begin{array}{cccc}\label{evolution}
e^{-i\frac{a+2}{2}\bar{t}}+e^{i(a-1)\bar{t}}&0&0&e^{i(a-1)\bar{t}}-e^{-i\frac{a+2}{2}\bar{t}}\\
0&1+e^{-i\frac{a-4}{2}\bar{t}}&e^{-i\frac{a-4}{2}\bar{t}}-1&0\\
0&e^{-i\frac{a-4}{2}\bar{t}}-1&1+e^{-i\frac{a-4}{2}\bar{t}}&0\\
e^{i(a-1)\bar{t}}-e^{-i\frac{a+2}{2}\bar{t}}&0&0&e^{-i\frac{a+2}{2}\bar{t}}+e^{i(a-1)\bar{t}}
\end{array}
\right),
\end{array}
\end{equation}
where the dimensionless time $\bar{t}=d\tau$.
Below we will use the dimensionless time in all calculations.
In particular, the evolution operator $U((a+2)\tau)$ will be written in terms of the dimensionless time $\bar{t}$.
The matrix (\ref{evolution}) is central-symmetric (CS) ($u_{i,j}=u_{5-i,5-j},\,i,j=1,2,3,4$) \cite{feldman2}. Using the orthogonal transformation
\begin{equation}\label{g}
G=\frac{1}{\sqrt{2}}
\left(
\begin{array}{cccc}
1&0&0&1\\
0&1&1&0\\
0&1&-1&0\\
1&0&0&-1
\end{array}
\right),
\end{equation}
one can decompose a CS-matrix into two blocks of size $2\times2$ ($G=G^{-1}$). However, in the considered case
\begin{equation}
GU((a+2)\tau)G=\left(
\begin{array}{cccc}
e^{i(a-1)\bar{t}}&0&0&0\\
0&e^{-i\frac{a-4}{2}\bar{t}}&0&0\\
0&0&1&0\\
0&0&0&e^{-i\frac{a+2}{2}\bar{t}}
\end{array}\right),
\end{equation}
the matrix of Eq.~(\ref{operator_matrix}) can be reduced to a diagonal form. As a result, we obtain the following analytic expression for the matrix representation of $\rho((a+2)M\tau)$ after $M$ periods of the pulse sequence of Eq.~(\ref{eq3}):
\begin{equation}\label{eq14}
\begin{array}{l}
\rho((a+2)M\tau)=U((a+2)M\tau)\rho_0U^+((a+2)M\tau)=\displaystyle{\frac{1}{8 \cosh^2\frac{\beta}{2}}}\nonumber\\
\bigskip
{}\times\left(
\begin{array}{cccc}
2 \cosh^2\frac{\beta}{2}  &  \sinh \beta e^{i\frac{3a-6}{2}M\bar{t}} &   \sinh\beta e^{i\frac{3a-6}{2}M\bar{t}}& 2 \sinh^2 \frac{\beta}{2}\\
\sinh \beta e^{-i\frac{3a-6}{2} M\bar t} &2 \cosh^2 \frac{\beta}{2} & 2 \sinh^2\frac{\beta}{2}&\sinh \beta e^{-i\frac{3a-6}{2}M\bar{t}}\\
\sinh \beta e^{-i\frac{3a-6}{2}M\bar{t}} & 2 \sinh^2 \frac{\beta}{2}&2 \cosh^2\frac{\beta}{2}& \sinh \beta e^{-i\frac{3a-6}{2}M\bar{t}}\\
2\sinh^2\frac{\beta}{2}&\sinh \beta e^{i\frac{3a-6}{2} M\bar{t}}&\sinh \beta e^{i\frac{3a-6}{2}M\bar{t}}&2\cosh^2\frac{\beta}{2}
\end{array}
\right)
\end{array}
\end{equation}
One can see from Eq.(\ref{eq14}) that the density matrix $\rho((a+2)M\tau)$ does not depend on time at $a=2$. Since the initial density matrix of Eq.~(\ref{dmatrix}) is separable, one can conclude that entanglement does not emerge in such conditions at arbitrary temperatures. Note that schemes $A$ and $B$ coincide at $a=2$ and $\varphi=\pi/2$, i.e., for $90^\circ$-pulses. This means that entanglement does not emerge in either scheme at such conditions even at long times. At the same time, entanglement can emerge when $a\neq 2$. 
A further investigation of entanglement for scheme $A$  is based on the Wootters approach \cite{hill,wootters}. 
According to the approach \cite{hill,wootters}, we should investigate the concurrence, which is determined via the square roots of the eigenvalues of the product of the matrices $\rho((a+2)M\tau)\tilde{\rho}((a+2)M\tau)$, where
\begin{equation}
\tilde{\rho}((a+2)M\tau)=(\sigma_y\otimes\sigma_y)\rho^\ast((a+2)M\tau)(\sigma_y\otimes\sigma_y))
\end{equation}
and $\sigma_y$ is the Pauli matrix.

Notice that the product of the matrices $\rho((a+2)M\tau) \tilde{\rho}((a+2)M\tau)$ is again a CS matrix.
The elements of that matrix are too lengthy and we give only its structure
\begin{equation}\label{structure}
\rho((a+2)M\tau)\tilde{\rho}((a+2)M\tau)=
\left(
\begin{array}{cccc}
p_1 & p_2+i p_3&p_4+ip_5&p_6\\
p_2-i p_3&\frac{1}{2}-p_1&p_7&p_4-i p_5\\
p_4-i p_5&p_7&\frac{1}{2}-p_1&p_2-i p_3\\
p_6&p_4+ip_5&p_2+ip_3&p_1
\end{array}
\right),
\end{equation}
where $p_i$ (i=1,2, \dots,7) are real parameters. After the orthogonal transformation (\ref{g}) the matrix (\ref{structure}) takes the block-diagonal form with $2\times 2$ subblocks. As a result, the analytic expressions for the square roots of eigenvalues can be obtained (see Appendix in Ref.~\cite{feldman2}). In our case, those values are \cite{feldman2}:
\begin{equation}\label{Eq16}
\begin{array}{c}
\lambda_1=\frac{1}{2}
\big\{
\frac{1}{2}\sqrt{(1+\tanh^2\frac{\beta}{2})^2-4\tanh^2\frac{\beta}{2}\cos^2\left(\frac{3a-6}{2}M\bar{t}\right)}\\
 \qquad\qquad\qquad\qquad\qquad\qquad\qquad\qquad
{}+\tanh\frac{\beta}{2}\left|\sin\left(\frac{3a-6}{2}M\bar{t}\right)\right|
\big\},
\end{array}
\end{equation}
\begin{equation}\label{Eq17}
\begin{array}{c}
\lambda_2=\frac{1}{2}
\big\{
\frac{1}{2}\sqrt{(1+\tanh^2\frac{\beta}{2})^2-4\tanh^2\frac{\beta}{2}\cos^2\left(\frac{3a-6}{2}M\bar{t}\right)}\\
 \qquad\qquad\qquad\qquad\qquad\qquad\qquad\qquad
{}-\tanh\frac{\beta}{2}\left|\sin\left(\frac{3a-6}{2}M\bar{t}\right)\right|
\big\},
\end{array}
\end{equation}
\begin{equation}\label{Eq18}
\lambda_3=\lambda_4=\frac{1-\tanh^2\frac{\beta}{2}}{4}.
\end{equation}
Using formulas (\ref{Eq16})-(\ref{Eq18}), we express the concurrence \cite{hill,wootters}
\begin{equation}\label{Eq19}
C=\max\left\{0,2\lambda_{\max}-\lambda_1-\lambda_2-\lambda_3-\lambda_4\right\},
\end{equation}
where $\lambda_{\max} =\max\left\{\lambda_1,\lambda_2,\lambda_3,\lambda_4\right\}$ as
\begin{equation}\label{Eq20}
C=\max\left\{0,\tanh\frac{\beta}{2}\left|\sin\left(\frac{3a-6}{2}M\bar{t}\right)\right|-\frac{1-\tanh^2\beta/2}{2}\right\}.
\end{equation}
Entanglement emerges at the critical temperature
\begin{equation}\label{Eq21}
T_{cr}=\frac{\hbar \omega_0}{k_B |\ln(\sqrt 2 -1)|},
\end{equation}
where $\hbar$ and $k_B$ are the Plank and Boltzmann constants, respectively ($\hbar=1.054\cdot10^{-34}\, \textup{m}^2 \textup{kg} / \textup{s}$, $k_B=1.38\cdot 10^{-23} \, \textup{m}^2 \textup{kg}\cdot \textup{s}^{-2} \textup{K}^{-1}$).
\begin{figure}
  \includegraphics[scale=0.3]{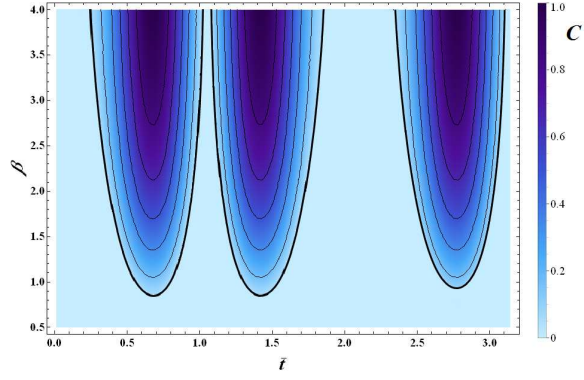}
\caption{The concurrence $C$ versus the dimensionless time $\bar{t}=d\tau$ and  the inverse temperature $\beta$ for Scheme $B$ of the multiple-pulse spin-locking experiment.The zeroth concurrence corresponds to separable states and  is shown in the light region. The regions of the entangled states and the separable ones are separated by the bold lines}
\label{fig:3}       % Give a unique label
\end{figure}
Entangled states exist at the temperatures less than $T_{cr}$.
If one takes $\omega_0=2\pi\cdot 500\times 10^6\;s^{-1}$,  entangled states emerge at the temperature $T_{cr}=27\;mK$.

The multiple-pulse spin locking experiments allow us to investigate experimentally emergence and development of quantum correlations in the considered system.
Indeed, one can find the magnetization $\mathfrak{M}_x((a+2)M\tau)$, which is observed in this experiment, using Eq.~(\ref{eq14}):
\begin{eqnarray}
\mathfrak{M}_x ((a+2)M\tau)=\gamma \hbar\left\langle I_x \right\rangle=\gamma\hbar\left\langle (I_{1x}+I_{2x})\right\rangle=\Tr \left\{ \rho((a+2)M\tau)I_x\right\}= \nonumber\\
\gamma\hbar \tanh {\frac{\beta}{2}}\cos\left(\frac{3a-6}{2} M\bar{t}\right).
\end{eqnarray}
One can express the concurrence $C$ of Eq.~(\ref{Eq20}) through the magnetization $\mathfrak{M}_x((a+2)M\tau)$:
\begin{equation}
C=\max\left\{0,\frac{1}{\gamma \hbar} \sqrt{(\gamma\hbar\tanh \frac{\beta}{2})^2-\mathfrak{M}_x^2((a+2)M\tau)}-\frac{1-\tanh^2\frac{\beta}{2}}{2}\right\}.
\end{equation}
Observing the magnetization $\mathfrak{M_x}((a+2)M\tau)$  in the course of the  experiment, we can obtain information about of the evolution of the quantum correlations in the system.

Crystallohydrates (lithium sulphate monohydrate $\mathrm{Li_2SO_4\ast H_2O}$, lithium chloride monohydrate $\mathrm{LiCl\ast H_2O}$, etc.) could be used for implementing the suggested experiments. Those crystallohydrates contain spin pairs ($s=1/2$) connected by the DDI. The pairs are far away from other proton spins. Thus, we indeed have isolated two-spin systems with the DDI in a strong external magnetic field.

We showed that entanglement is absent in scheme $B$ of the multiple pulse spin locking (Fig.~\ref{fig:2}) for $\pi/2$-pulses.
The situation is essentially  different, if $\varphi\neq\pi/2$.
For example, at $\varphi =\pi/4$ the Hamiltonian matrix assumes four different values $H_{dz}$, $H_{\pi/4}$, $H_{dy}$, $H_{3\pi/4}$ on the period of the sequence (\ref{Eq4}).
\begin{equation}\label{Eq22}
H_{\pi/4}=e^{-i\frac{\pi}{4} I_x}H_{dz}e^{i\frac{\pi}{4}I_x}=\frac{d}{8}
\left(
\begin{array}{cccc}
1&-3i&-3i&-3\\
3i&-1&-1&3i\\
3i&-1&-1&3i\\
-3&-3i&-3i&1
\end{array}
\right);
\end{equation}

\begin{eqnarray}
H_{3\pi/4}=e^{-\frac{3\pi}{4}iI_x} H_{dz} e ^{i\frac{3\pi}{4}I_x}=
e^{-i\frac{3\pi}{4}I_x} e^{i\pi I_x} H_{dz} e^{-i\pi I_x} e^{i\frac{3\pi}{4}I_x}=\\ \nonumber
\qquad e^{i\frac{\pi}{4}I_x}H_{dz}e^{-i\frac{\pi}{4} I_x}=
\frac{d}{8}
\left(
\begin{array}{cccc}
1&3i&3i&-3\\
-3i&-1&-1&-3i\\
-3i&-1&-1&-3i\\
-3&3i&3i&1
\end{array}
\right),
\end{eqnarray}

where we take into account that $\left[H_{dz}, e^{i\pi I_x}\right]=0$ and
 \begin{equation}
e^{-i\frac{\pi}{4}I_x}I_{iz}e^{i\frac{\pi}{4}I_x}=e^{-i\frac{\pi}{4} I_{ix}}I_{iz}e^{i\frac{\pi}{4}I_{ix}}=\frac{1}{\sqrt{2}}(I_{iz}-I_{iy}),\quad i=1,2.
\end{equation}
The evolution operator $U(8\tau)$ on the period  of the sequence (\ref{Eq4}) is
\begin{eqnarray}\label{Eq23} 
U(8\tau)=\exp(-i\tau H_{dz})\cdot\exp(-i2\tau H_{3\pi/4})&\cdot& \exp(-i2\tau H_{dy})\\
&\cdot&\exp(-i2\tau H_{\pi/4}) \cdot\exp(-i\tau H_{dz}).\nonumber
\end{eqnarray}
Taking into account that the evolution operator of Eq.~(\ref{Eq23}) is CS  \cite{feldman2}, one can find  the evolution operator $U(8M\tau)$  after $M$ periods of the sequence (\ref{Eq4}).
The expression for the corresponding density matrix
\begin{equation}
\rho(8M\tau)=U(8M\tau)\rho_0 U^+(8M\tau)
\end{equation}
is very long and we omit it.
We restrict ourselves to a set of eigenvalues, analogous to Eqs.~(\ref{Eq16}), (\ref{Eq17}), (\ref{Eq18}), which are needed for the calculation of the concurrence (see Eq. (\ref{Eq19})).
These eigenvalues are given by the expressions \cite{feldman2}:
\begin{equation}
\lambda_1=\frac{1}{2}\left\{
\sqrt{(2p_1+p_6-\frac{1}{2}-p_7)^2+16p_3^2}+\sqrt{\left(\frac{1}{2}+p_6+p_7\right)^2-16p_4^2}
\right\},
\end{equation}
\begin{equation}
\lambda_2=\frac{1}{2}\left\{
\sqrt{(2p_1+p_6-\frac{1}{2}-p_7)^2+16p_3^2}-\sqrt{\left(\frac{1}{2}+p_6+p_7\right)^2-16p_4^2}
\right\},
\end{equation}
\begin{equation}
\lambda_3=\frac{1}{2}
\left\{
\left|2p_1-p_6-\frac{1}{2}+p_7\right|+\left|\frac{1}{2}-p_6-p_7\right|
\right\},
\end{equation}
\begin{equation}
\lambda_4=\frac{1}{2}
\left|
\left|2p_1-p_6-\frac{1}{2}+p_7\right|-\left|\frac{1}{2}-p_6-p_7\right|
\right|.
\end{equation}
The parameters $p_1$, $p_3$, $p_4$, $p_6$, $p_7$ are determined as
\begin{equation}
p_1=\frac{1}{4}+\frac{\sin^2(M\Phi(\bar{t}))}{q^2(\bar{t})}s(\bar{t})\cdot r(\bar{t}) \tanh \frac{\beta}{2},
\end{equation}
\begin{equation}
p_3=\frac{1}{4}\sin(2M\Phi(\bar{t}))\frac{s(\bar{t})}{q(\bar{t})}\tanh\frac{\beta}{2},
\end{equation}
\begin{equation}
p_4=\frac{1}{4}\left\{1-\frac{2\sin^2(M\Phi(\bar{t}))}{q^2(\bar{t})}s^2(\bar{t})
\right\}
\tanh\frac{\beta}{2},
\end{equation}
\begin{equation}
p_6=\frac{1}{4}\left\{
\tanh^2\frac{\beta}{2}+\frac{4\sin^2(M\Phi(\bar{t}))}{q^2(\bar{t})}s(\bar{t})\cdot r(\bar{t}) \tanh\frac{\beta}{2}
\right\},
\end{equation}
\begin{equation}
p_7=\frac{1}{4}\left\{
\tanh^2\frac{\beta}{2}-\frac{4\sin^2(M\Phi(\bar{t}))}{q^2(\bar{t})}s(\bar{t})\cdot r(\bar{t})\tanh \frac{\beta}{2}
\right\},
\end{equation}
where
\begin{equation}
s(\bar{t})=\sin(6\bar{t})-2\sin(3\bar{t}), 
\qquad
r(\bar{t})=\cos\left(\frac{9\bar{t}}{2}\right)-\cos\left(\frac{3\bar{t}}{2}\right),
\end{equation}
\begin{equation}
q(\bar{t})=\frac{1}{2}\sqrt{26-24\cos(6\bar{t})-2\cos(12\bar{t})+16\cos(9\bar{t})-16\cos(3\bar{t})},
\end{equation}
and
\begin{equation}
\Phi(\bar{t})=\arctan\left(\frac{q(\bar{t})}{p(\bar{t})}\right),\qquad p(\bar{t})=4\cos(3\bar{t})-\cos(6\bar{t})+1.
\end{equation}
The concurrence can be computed from the above equations and Eq.~(\ref{Eq19}).
The concurrence is presented as a function of two parameters: the dimensionless time $\bar{t}$ and the inverse temperature $\beta$ in Fig.~\ref{fig:3} for $M=1$.
Time intervals with nonzero entanglement alternate with time intervals during which the state is separable.
The concurrence increases when the temperature decreases.
The time intervals of nonzero concurrence grow when the temperature decreases.
At smaller inverse temperatures $\beta$, a greater number $M$ of periods is required for the emergence of entanglement. The growth of the temperature reduces the entanglement area. For example, $M=8$ periods are necessary for the emergence of entanglement at the inverse temperature $\beta=3$  at $\varphi=\pi/4$, $\bar{t}=0.1$.
At $\beta\geq 5$, entanglement emerges at any number of the periods.

Notice that $\beta=3$ corresponds to the temperature $T=\frac{\hbar \omega_0}{k_B \beta}\approx 8 \textup{mK}$ at $\omega_0=2\pi\cdot 500\cdot 10^6 \,\textup{s}^{-1}$ (see Eq.~(\ref{dmatrix})), and $\beta=5$ corresponds to $T=4.8 \textup{mK}$.
Such temperatures can be obtained by using the method of adiabatic demagnetization in a rotating reference frame \cite{B1}. Note that even microKelvin temperatures were  obtained for investigations of the dipolar magnetic ordering \cite{B1}.

\section{Quantum discord in the multiple-pulse spin-locking NMR experiments}
Quantum correlations in many-particle systems are responsible for the effective work of quantum devices (in particular quantum computers) and give  significant advantages over their classical counterparts.
However, quantum correlations are determined by entanglement only for pure states.
It turned out that quantum algorithms \cite{A1} which significantly  outperform the classical counterparts can work using mixed separable (non-entangled) states.
Furthermore, it turned out that quantum non-locality can be observed in some separable states without entanglement \cite{A2,ref19,ref20}. 
According to the current understanding \cite{ollivier,henderson}, total (quantum and classical) correlations can be determined with the quantum conditional entropy \cite{henderson} which can be obtained after performing a complete set of projective measurements carried out  only  over one of the subsystems. 
Then a measure of quantum correlations (the quantum discord) is determined as the difference between the mutual information and its classical part minimized over all possible projective measurements \cite{ollivier,henderson}. 
The quantum discord is determined completely by the quantum properties of the system, coincides with entanglement for pure states \cite{A3}, and equals zero for classical systems.

Here we investigate quantum discord in multiple-pulse NMR spin-locking experiments.
First, we calculate analytically quantum discord in the high temperature approximation \cite{goldman} for scheme $A$. 
Then we investigate quantum discord in scheme $A$ at arbitrary temperatures.

\subsection{Quantum discord in scheme $A$ of  multiple-pulse spin-locking at high temperatures}
We write the density matrix $\rho((a+2)M\tau)$ of Eq.~(\ref{eq14}) for scheme $A$ of multiple-pulse spin-locking in the Bloch diagonal form  in the high temperature approximation \cite{goldman} for the inverse temperature $\beta \ll 1$:
\begin{equation}\label{Eq38}
\rho((a+2)M\tau)=\frac{1}{4}I\otimes I+\frac{A_{1x}}{4}(\sigma_x\otimes I+I\otimes\sigma_x)+\frac{A_{zy}}{4}(\sigma_y\otimes \sigma_z+\sigma_z \otimes \sigma_y),
\end{equation}
where $I$ is the identity operator, $\sigma_\alpha$ ($\alpha=x,y,z$) are the Pauli matrices, and
\begin{equation}\label{Eq39}
A_{1x}=\frac{\beta}{2}\cos(x),\quad A_{zy}=-\frac{\beta}{2}\sin (x),\quad x=\frac{3}{2}(a-2)M\bar{t}.
\end{equation}

According to the standard approach \cite{ref21,ref22} for the optimization of the quantum conditional entropy, one performs a total set of projective measurements over one-qubit subsystem $Q_k=V\Pi_k V^+$,
where the matrix $V\in SU(2)$ and $\Pi_k$ ($k=0,1$) are projectors.
It is suitable to write the projectors via the Pauli matrices \cite{ref23}:
\begin{equation}\label{Eq40}
\Pi_0=\frac{1}{2} (I+n_x \sigma_x +n_y \sigma_y +n_z \sigma_z), \quad \Pi_1=\frac{1}{2}(I-n_x \sigma_x -n_y \sigma_y -n_z \sigma_z),
\end{equation}
where $n_x$, $n_y$, $n_z$ are the components of the unit vector ($n_x^2+n_y^2+n_z^2=1$).

The density matrix $\rho((a+2)M\tau)$ of Eq.~(\ref{Eq38}) can be transformed after performing measurements, described by the projectors of Eq.~(\ref{Eq40}), as
\begin{equation}\label{Eq41}
\Pi_k\rho((a+2)M\tau)\Pi_k=\frac{1}{4}
\left[
I+A_{1x}(\sigma_x\pm n_x I)\pm A_{zy} (n_y \sigma_z +n_z \sigma_y)
\right]
\otimes \Pi_k.
\end{equation}
Thus, one can find that the whole system is described by the ensemble of the states $\{p_k,\rho_k\}$ ($k=0,1$) after the measurements, where
\begin{eqnarray}\label{Eq42}
p_0=\Tr\left\{
(I\otimes \Pi_0)\rho((a+2)M\tau)(I\otimes\Pi_0)
\right\}=
\frac{1}{2}(1+A_{1x}n_x),\label{Eq:42}\\ \nonumber
p_1=\Tr\left\{
(I\otimes \Pi_1) \rho((a+2)M\tau)(I\otimes \Pi_1)
\right\}=
\frac{1}{2}(1-A_{1x}n_x),
\end{eqnarray}
and the matrices $\rho_0$, $\rho_1$ are
\begin{eqnarray}\label{Eq:43}
\rho_0=\frac{1}{4p_0} \left[
I+A_{1x}(n_x I +\sigma_x)+A_{zy}(n_y \sigma_z +n_x\sigma_y)
\right]\otimes \Pi_0,  \\ \nonumber
\rho_1=\frac{1}{4p_1}\left[ I+A_{1x}(\sigma_x - n_x I) - A_{zy} (n_y \sigma_z +n_z \sigma_y)\right]\otimes \Pi_1.
\end{eqnarray}
Then the conditional quantum entropy $S_{cond}$ after the measurements over the second subsystem of the two-spin system can be written \cite{ref21,ref22} as
\begin{equation}\label{Eq43}
S_{cond}=p_0 S(\rho_0) + p_1 S(\rho_1),
\end{equation}
where $S(\rho_k)=-\Tr[\rho_k \log_2 \rho_k]$.
The calculation of the quantum conditional entropy (\ref{Eq43}) with Eqs.~(\ref{Eq42}), (\ref{Eq:43}) up to the terms of the order $\beta^2$ leads to the formula
\begin{equation}\label{Eq44}
S_{cond}=1-\frac{1}{2\ln 2}(A_{1x}^2+A_{zy}^2(1-n_x^2)).
\end{equation}
The quantum conditional entropy of Eq.~(\ref{Eq44}) achieves its minimal value at $n_x=0$ ($n_z^2+n_y^2=1$). 
To calculate quantum discord, we will also use the expression for the entropy $S(\rho)$ of the system
\begin{equation}\label{Eq45}
S(\rho)=2-\frac{\beta^2}{4\ln2}.
\end{equation}
The expressions for the entropies $S(\rho_A)$ and $S(\rho_B)$ of the first and second subsystems are
\begin{equation}\label{Eq46}
S(\rho_A)=S(\rho_B)=1-\frac{\beta^2\cos^2 x}{8\ln 2}.
\end{equation}
We used the reduced density matrices $\rho_A$ and $\rho_B$ for the calculations in Eq.~(\ref{Eq46})
\begin{eqnarray}
\rho_A=\Tr_B\left\{\rho(t)\right\}=\frac{1}{2}I+\frac{\beta}{4}\cos{x}\cdot\sigma_{1x}, \nonumber\\ 
\rho_B=\Tr_A\left\{\rho(t)\right\}=\frac{1}{2}I+\frac{\beta}{4}\cos{x}\cdot \sigma_{2x}.
\end{eqnarray}
Quantum discord $D$ is the difference of the two definitions of the mutual quantum information \cite{ollivier,henderson}. Discord can be written in our case as
\begin{equation}\label{discord}
D=S(\rho_B)-S(\rho)+S^{min}_{cond},
\end{equation}
where $S^{min}_{cond}$ is the minimal value of the quantum conditional entropy $S_{cond}$ (see Eq.~(\ref{Eq44}) at $n_x=0$).
Using Eqs.~(\ref{Eq44}), (\ref{Eq45}), (\ref{Eq46}), (\ref{discord}) one can calculate the corresponding quantum discord
\begin{equation}\label{Eq48}
D=\frac{\beta^2}{8\ln 2}\sin^2{x}=\frac{\beta^2}{8 \ln 2} \sin^2 \frac{3}{2}((a-2)M\bar{t}).
\end{equation}
One can see from Eq.~(\ref{Eq48}) that $D=0$ at $a=2$.
This means that in the multiple-pulse spin-locking NMR experiment with $90^\circ$-radio-frequency pulses, quantum discord equals zero, as entanglement does (see Section~3). 
Quantum correlations are absent in this case, both for Scheme A and Scheme B.
We emphasize that quantum discord is obtained analytically for scheme $A$ of the multiple-pulse spin-locking in the high temperature approximation.

\subsection{Quantum discord in scheme $A$ of multiple-pulse spin-locking at arbitrary temperatures}

We transform the density matrix $\rho((a+2)M\tau)$ of Eq.~(\ref{eq14}) with the local Hadamard transformation R \cite{yurNMR}
\begin{equation}
R=\frac{1}{\sqrt{2}}
\left(
\begin{array}{cc}
1&1\\1&-1
\end{array}
\right)
\otimes
\left(
\begin{array}{cc}
1&1\\
1&-1
\end{array}
\right)=
\frac{1}{2}\left(
\begin{array}{cccc}
1&1&1&1\\
1&-1&1&-1\\
1&1&-1&-1\\
1&-1&-1&1
\end{array}\right).
\end{equation}
The transformed density matrix is
%\begin{equation}
   \begin{eqnarray}
R\rho((a+2)M\tau)R^+=\qquad\qquad\qquad\qquad\qquad\qquad \label{eqn58}\\
\frac{1}{2}\left(
\begin{array}{cccc}
\frac{1}{2}+\cos(x)\tanh\frac{\beta}{2}+\frac{1}{2}\tanh^2\frac{\beta}{2}&0&0&-i\sin(x)\tanh\frac{\beta}{2}\\
0&\frac{1}{2}(1-\tanh^2\frac{\beta}{2})&0&0\\
0&0&\frac{1}{2}(1-\tanh^2\frac{\beta}{2})&0\\
i\sin(x)\cdot\tanh\frac{\beta}{2}&0&0&\frac{1}{2}-\cos(x)\tanh\frac{\beta}{2}+\frac{1}{2}\tanh^2\frac{\beta}{2}
\end{array}
\right).\nonumber
\end{eqnarray}
%\end{equation}
The density matrix of Eq.~(\ref{eqn58}) is a matrix of an $X$ state \cite{ref22}. A method of the calculation of the quantum discord for $X$ states is developed in \cite{ref22}. Since the local transformations do not change the quantum discord, one can apply the method \cite{ref22} for the considered case.
Instead of the projectors of Eq.~(\ref{Eq40}) we will use the following \cite{fanchini,ciliberti}
\begin{equation}
\Pi_0=\left(
\begin{array}{cc}
\cos^2{\theta/2}&\frac{1}{2}e^{-i\Phi}\sin{\theta}\\
\frac{1}{2}e^{i\Phi}\sin{\theta}&\sin^2{\theta/2}
\end{array}
\right),
\quad
\Pi_1=\left(
\begin{array}{cc}
\sin^2{\theta/2}&-\frac{1}{2}e^{-i\Phi}\sin{\theta}\\
-\frac{1}{2}e^{i\Phi}\sin\theta&\cos^2\theta/2,
\end{array}
\right)
\end{equation}
where $\theta$, $\Phi$ are spherical coordinates.
It was shown \cite{ciliberti} that the minimal conditional entropy is achieved at $\cos(2\Phi)=1$.
As a result, all further calculations depend only on one parameter $\theta$.
According to Ref.~\cite{yurNMR}, quantum discord $D$ can be expressed as 
\begin{equation}
D=\min(D_0,D_\theta,D_{\pi/2}),
\end{equation}
where $D_0$ and $D_{\pi/2}$ are the boundary values of quantum discord at $\theta=0,\pi/2$ and $D_\theta$ is discord at the given value of $\theta$.
The reduced density matrices of the two subsystems of the two-spin system are
\begin{equation}
\rho_A=\rho_B=
\left(\begin{array}{cc}
\bar{a}+b&0\\
0&b+d
\end{array}
\right),
\end{equation}
where
\begin{eqnarray}
\bar{a}&=&\frac{1}{4}+\frac{1}{2}\tanh\frac{\beta}{2}\cdot\cos(x)+\frac{1}{4}\tanh^2\frac{\beta}{2},\nonumber \\
b&=&\frac{1}{4}-\frac{1}{4}\tanh^2\frac{\beta}{2},\label{Eq54}\\ \nonumber
d&=&\frac{1}{4}-\frac{1}{2}\tanh\frac{\beta}{2}\cdot\cos(x)+\frac{1}{4}\tanh^2\frac{\beta}{2}.
\end{eqnarray}
The entropies $S(\rho_A)$, $S(\rho_B)$ of the subsystems are
\begin{equation}\label{Eq55}
S(\rho_A)=S(\rho_B)=-(\bar{a}+b)\log_2(\bar{a}+b)-(b+d)\log_2(b+d),
\end{equation}
where $\bar{a}$, $b$, $d$ are determined by Eq.~(\ref{Eq54}).
The entropy  of the total system is
\begin{eqnarray}\label{Eq56}
S(\rho)&=&-\frac{\bar{a}+d+\sqrt{(\bar{a}-d)^2+4u^2}}{2}\log_2 \left( \frac{\bar{a}+d+\sqrt{(\bar{a}-d)^2+4u^2}}{2}\right)\nonumber\\ {}&-&
\frac{\bar{a}+d-\sqrt{(\bar{a}-d)^2+4u^2}}{2}\log_2 \left( \frac{\bar{a}+d-\sqrt{(\bar{a}-d)^2+4u^2}}{2}\right) \nonumber\\&&{}-2 b\log_2 b,
\end{eqnarray}
where
\begin{equation}\label{Eq57}
u=-\frac{1}{2}\tanh\left( \frac{\beta}{2}\right)\sin(x).
\end{equation}
\begin{figure}[ht]
  \includegraphics[scale=0.3]{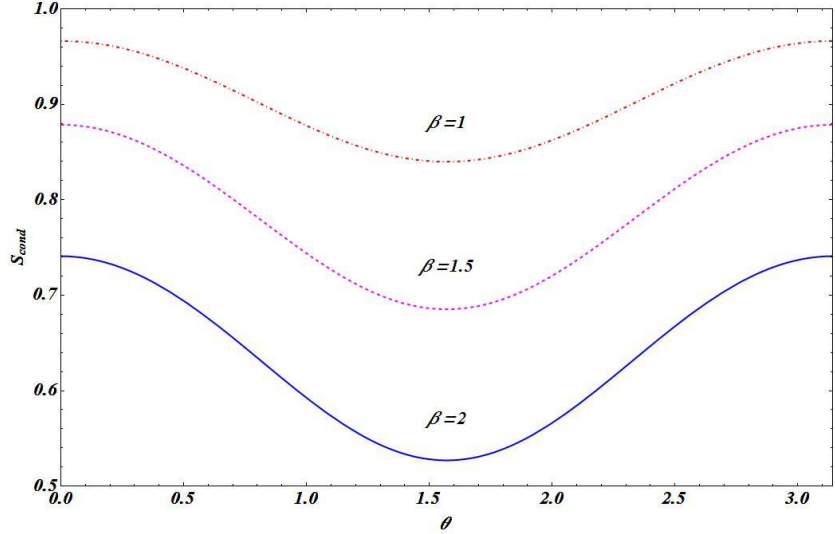} 
\caption{Dependence of quantum conditional entropy $S_{cond}$ on the spherical coordinate $\theta$ at different inverse temperatures $\beta=1,\,1.5$ and 2 at $\bar{t}=1$, $M=1$, $a=3$.}
\label{fig:4}       % Give a unique label
\end{figure}
In order to find the minimal value of the quantum conditional entropy at different $x$ and $\beta$ ( $0<x<150$, $0<\beta<7$) we used the random mutation algorithm \cite{C1} which is a modification of a simplified genetic optimization.
In all cases, we have found that the minimal value is obtained at $\theta=\pi/2$. Fig.~\ref{fig:4} demonstrates that the minimal value of quantum conditional entropy is indeed achieved at $\theta=\pi/2$ at different temperatures in the course of the system evolution in the multiple-pulse spin locking experiment.

\begin{figure}[ht]
\begin{tabular}{cc}
  \includegraphics[scale=0.13]{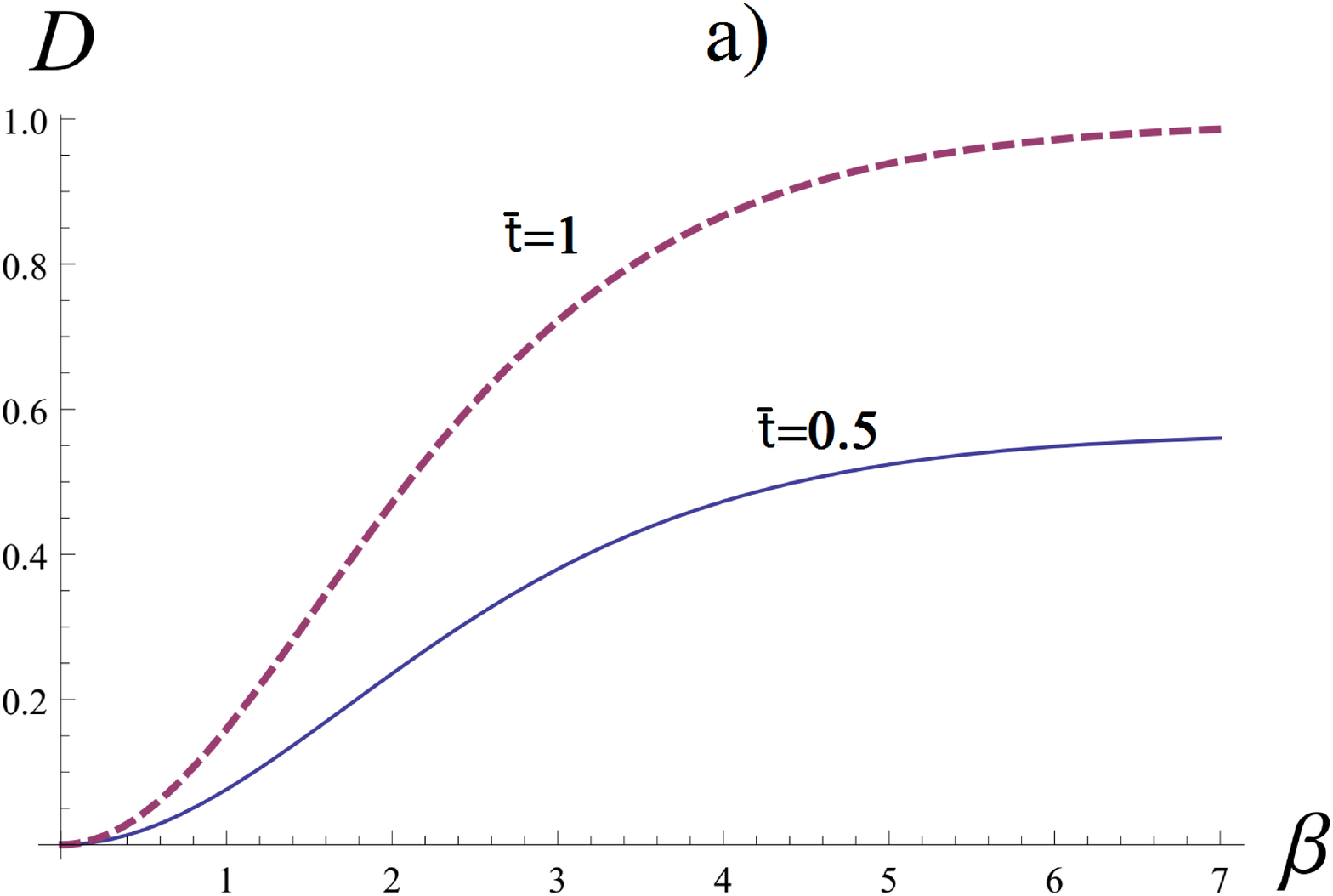} 
	&
	\includegraphics[scale=0.13]{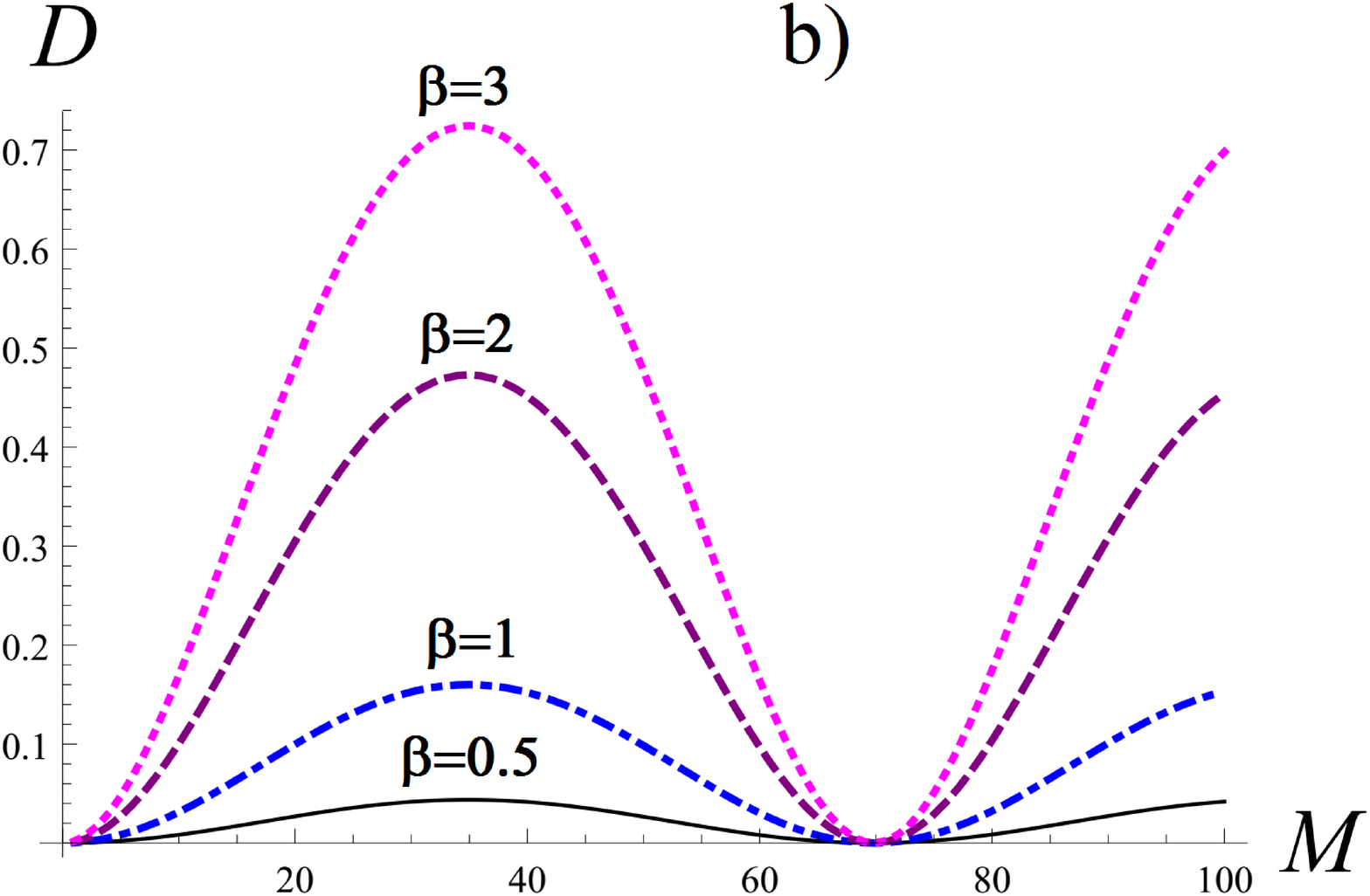}
	\end{tabular}
\caption{Dependence of the quantum discord on the inverse temperature $\beta$ at $\bar{t}=0.5$ and $\bar{t}=1$; $M=1$, $a=3$ (Fig.~5a). Dependence of the quantum discord on the number of periods of the pulse sequence (Fig.~1) at $\bar{t}=0.003$ and $\beta=0.5$, 1, 2, and 3 (Fig.~5b). }
\label{fig:5}       % Give a unique label
\end{figure}

 The minimal value of the quantum conditional entropy $S_{cond}$ is
\begin{eqnarray}
S_{cond}=-\frac{1+\sqrt{(\bar{a}-d)^2+4u^2}}{2}\log_2 \frac{1+\sqrt{(\bar{a}-d)^2+4u^2}}{2} \\
- \frac{1-\sqrt{(\bar{a}-d)^2+4u^2}}{2}\log_2 \frac{1-\sqrt{(\bar{a}-d)^2+4u^2}}{2} \nonumber
\end{eqnarray}
Using this result and Eqs.~(\ref{Eq55}), (\ref{Eq56}), (\ref{Eq57}), one can obtain that quantum discord is 
\begin{eqnarray}\label{Eq58}
D&=&D_{\pi/2}=-S(\rho)-(\bar{a}+b)\log_2(\bar{a}+b)-(b+d)\log_2(b+d)\\ \nonumber
{}&-&\left( \frac{1+\sqrt{(\bar{a}-d)^2+4u^2}}{2}\right)\log_2\left( \frac{1+\sqrt{(\bar{a}-d)^2+4u^2}}{2}\right)\\
\nonumber &&{}-\left(\frac{ 1-\sqrt{(\bar{a}-d)^2+4u^2}}{2}\right)\log_2\left( \frac{1-\sqrt{(\bar{a}-d)^2+4u^2}}{2}\right).
\end{eqnarray}
At high temperatures ($\beta\ll 1$), the result~(\ref{Eq58}) coincides with the one (\ref{Eq48}), obtained analytically.
As follows from Eq.~(\ref{Eq58}), quantum discord is absent at $a=2$.
This means that quantum correlations do not emerge in the multiple-pulse spin-locking experiment with $90^\circ$-pulses.
The dependencies of quantum discord on the inverse temperature and the number of periods of the multiple-pulse spin-locking experiment are given in Fig. \ref{fig:5} at $a=3$.
\section{Conclusion}
The multiple-pulse spin-locking experiment is an effective method for investigations of slow motions in solids.
At the same time this method is very instructive for the study of quantum correlations.
It is very important that milliKelvin temperatures required for an emergence of quantum correlations can be achieved in solid state NMR experiments \cite{B1}.

We calculated entanglement and quantum discord in two schemes of such an experiment.
Entanglement and quantum discord do not emerge in multiple-pulse spin-locking with $90^\circ$ -pulses at arbitrary temperatures. 
In some cases, quantum correlations emerge only after several periods of the irradiation of the spin system by a sequence of  r.f.pulses.
Quantum correlations increase when the temperature decreases.

The work is supported by Russian Foundation of Basic Research (Grant No.~16-03-00056) and the Program of the Presidium of the Russian Academy of Sciences No. 5 ''Electron spin resonance, spin-dependent electron effects and spin technologies''.

\end{document}